\RequirePackage{ifpdf}
\documentclass[12pt,letterpaper]{JHEP3}
\usepackage[english]{babel}
\usepackage{amsmath}
\usepackage{commath}
\usepackage{ textcomp }
\usepackage[normalem]{ulem}

\usepackage{cite}

\usepackage{soul}

\title{Gauge dependence of the AdS instability problem.}

\author{
Fotios V. Dimitrakopoulos\footnote{f.dimitrakopoulos@uva.nl},
Ben Freivogel\footnote{benfreivogel@gmail.com},
Juan F. Pedraza\footnote{jpedraza@uva.nl } 
and I-Sheng Yang\footnote{isheng.yang@gmail.com}\\
$^{*\dagger\ddagger}$ GRAPPA and ITFA, Institute of Physics, Universiteit van Amsterdam, \\
Science Park 904, 1090 GL Amsterdam, Netherlands \\ \ \\
$^\S$ Perimeter Institute of Theoretical Physics, \\
31 Caroline Street North, Waterloo, ON N2L 2Y5, Canada, \\ and, \\
Canadian Center of Theoretical Astrophysics, \\
60 St George St, Toronto, ON M5S 3H8, Canada
}

\abstract{Previous work on the AdS instability problem within the two-time framework (TTF) has found an ``oscillating singularity" whose presence depends on the gauge choice. We give a physical interpretation of this singularity as a diverging redshift between the boundary and the center of AdS. This signals  a genuine breakdown of the linearized gravity. One can also identify the diverging redshift through a back-reaction calculation purely in the boundary gauge, where the TTF result stays regular.

}

\begin{document}

\section{Introduction}
\label{sec:introduction}

The question of whether global $AdS$ spacetime is generically stable or unstable under small perturbations is a very interesting problem. A conclusive resolution still eludes us despite the combined efforts of many people\cite{BizMal15,Deppe16,CraEvn15a,BizRos11,MalRos13a,GreMai15,CraEvn14,BucGre14,DiaHor11,DiaHor12,DiaSan16,DFLY14,DimYan15,FreYan15,OliSop16,BalBuc14,CraEvn14a}.
The two-time framework (TTF) is a well-established tool that reduces the full gravitational dynamics into the ``slow-time'' evolution of complex amplitudes of approximate eigenstates. It operates on two approximations:
\begin{itemize}
\item The deviation from empty $AdS$ metric is small, so we can keep only the leading order gravitational back-reaction.
\item The evolution can be averaged over a ``fast'' time scale set by the $AdS$ radius, reducing to the dynamics in a ``slow'' time scale.
\end{itemize} 
One can simply follow the two-time evolution and observe whether the first approximation breaks down. If it does not, then the metric stays near empty $AdS$ and an instability is not triggered. If it does break down, then it implies an order one deviation from empty $AdS$, thus triggering an instability.

In \cite{BizMal15}, numerical results suggested that gravitational instability seems to coincide with a breakdown of TTF from an oscillating singularity---the complex amplitudes all start to acquire infinite phases. However, a direct logical link between the two was missing, because the physical interpretation of the oscillating singularity remained unclear. That is because a breakdown of TTF could be due to failure of either one of the two approximations, but only the breakdown of the first approximation has direct implications for the instability.\footnote{Some may have the intuition that the breakdown of the second approximation also can only come from large deviations from the $AdS$ metric, but such statement is never proven explicitly.} Later, in \cite{CraEvn15a}, was suggested that TTF might not suffer from an oscillating singularity if one chooses a different gauge, a fact that was subsequently verified numerically in \cite{Deppe16}. Those results appeared to add more confusion.

In this short note, we point out that the combination of \cite{BizMal15} and \cite{CraEvn15a,Deppe16} actually eliminates the confusion. A diverging difference\footnote{We will specify what this means in the subsequent sections.} between the results in two different gauges implies a diverging redshift between two different locations in $AdS$, which in turn implies a diverging deviation in the metric.
Alternatively, one could have used only the result in the boundary gauge where the TTF solutions stays finite \cite{Deppe16}.  Explicitly calculating its geometric back-reaction demonstrates the same divergence \cite{FreYan15}. 

Note that the actual geometric back-reaction is the TTF result multiplied by the amplitude squared of the initial perturbation. A diverging TTF redshift means that linearized gravity breaks down for arbitrarily small initial amplitude, which triggers a genuine instability of global $AdS$.

In section \ref{sec-Model}, we will briefly review the model of spherically symmetric scalar perturbations in $AdS$ and the perturbation theory leading to the Two Time Framework. In section \ref{sec-gauges}, we will compare the result between two different gauges. We will show that a diverging difference of the phases in these two different gauges is equivalent to a diverging ``averaged'' redshift calculated from back-reaction. When the averaged redshift diverges, the actual redshift must diverge at some moment, guaranteeing a large deviation from the empty $AdS$ metric.


\section{Review of the model}
\label{sec-Model}

The model that is mainly used is a perturbation in the form of a spherically symmetric massless scalar field which propagates under its own self--gravitation in the $AdS$ background. For the metric of asymptotically $AdS$ spacetimes we use the following ansatz\footnote{For simplicity we have set the $AdS$ radius to 1.}
\begin{equation}\label{eq:aAdSmetric}
ds^2 = \frac{1}{\text{cos}^2x}\left( -A e^{-2\delta} dt^2 + A^{-1}dx^2 + \text{sin}^2x d\Omega ^2 \right),
\end{equation}
where the functions $A$ and $\delta$ depend only on time $t$ and the radial coordinate $x \in \left[ 0, \frac{\pi}{2} \right] $. The metric (\ref{eq:aAdSmetric}) is not entirely gauge fixed. 
Two gauge fixing conditions that are common in the literature are: $\delta (t,0) = 0$ and $\delta (t,\pi / 2) = 0$. The first one constitutes the so called {\it central gauge}, and $t$ corresponds to the proper time in the center of $AdS$, while the second choice constitutes the {\it boundary gauge}, and $t$ corresponds to the proper time at the boundary.

The equations that govern the evolution of the system are the wave equation for a massless scalar field and the Einstein equations with a stress-energy tensor due to $\phi$. Using the variables  $\Phi = \phi ' $ and $\Pi = A^{-1}e^{\delta} \dot{\phi}$, the equations of motion can be written as
\begin{eqnarray}\label{eq:EOM1}
\dot{\Phi} =  \left( A e^{- \delta} \Pi \right) ' , \qquad \dot{\Pi}  =   \frac{1}{\text{tan}^x} \left( \text{tan}^2 x A e^{-\delta} \Phi \right) ' ,
\end{eqnarray}
while the Einstein equations reduce to the constraints
\begin{eqnarray}\label{eq:EOM2}
A' & = & \frac{1 + 2 \text{sin}^2x}{\text{sin}x\text{cos}x} \left( 1- A \right) - \text{sin}x \text{cos}x A \left( \Phi ^2 + \Pi ^2 \right) \nonumber \\
\delta ' & = & - \text{sin}x \text{cos}x \left( \Phi ^2 + \Pi ^2 \right).
\end{eqnarray}

We usually turn to perturbation theory to solve this system of equations. We start with some initial data of the form $\left( \phi, \dot{\phi} \right)_{t=0} = \left( \epsilon f(x), \epsilon g(x) \right)$ and we look for an approximate solution as a perturbative, in the amplitude $\epsilon$, expansion :
\begin{eqnarray}\label{eq:expansion}
\phi (t,x) = \sum_{k=0}^{\infty} \phi _{2k+1}(t,x) \epsilon ^{2k +1} , \quad A(t,x) = 1 + \sum _{k=1}^{\infty} A_{2k}(t,x) \epsilon ^{2k }, \quad \delta(t,x) = \sum _{k=1}^{\infty} \delta _{2k}(t,x) \epsilon ^{2k }. \nonumber \\
\end{eqnarray}
Inserting this ansatz into the equations of motion and collecting terms of the same order of $\epsilon$ we obtain a set of linear equations which can be solved order by order. 

To first order, we merely have a scalar filed propagating in the $AdS$ background
\begin{eqnarray}\label{eq:EOMphi1}
\ddot{\phi}_{1} + L \phi _1 = 0.
\end{eqnarray}
Here, $L = - \frac{1}{\text{tan}^{d-1} x} \partial _x \left( \text{tan}^{d-1} x \partial _x \right)$ is the Laplacian of $AdS_{d+1}$ with eigenvalues $\omega _j ^2 = (2j +d)^2$ and eigenfunctions
\begin{eqnarray}
e_j = d_j \text{cos}^d x P_j^{\left(\frac{d}{2}-1,\frac{d}{2}\right)}\left(\cos (2x)\right), \qquad d_j =\frac{2\sqrt{j!(j+d-1)!}}{\Gamma\left(j+\frac{d}{2}\right)}.
\end{eqnarray}
Solving eq.(\ref{eq:EOMphi1}) one simply gets:

\begin{eqnarray}
\phi _1 (t,x)  =  \sum_j c_j ^{(1)}(t) e_j(x) = \sum_{j} \left( \alpha _j e^{i \omega _j t} + \bar{\alpha}_j e^{-i \omega _j t} \right) e_j (x).\label{eq:phi1expr}
\end{eqnarray}

To second order we have the back--reaction in the metric described by $A_2$ and $\delta _2$. The solutions are:
\begin{eqnarray}
A_2(t,x) & = & -\nu (x) \int _{0}^{x} \left( \dot{\phi} _1 (t,y) ^2 + \phi ^{\prime} (t,y) ^2 \right) \mu (y) dy, \\
\delta _2 (t,x) & = & \begin{cases}
                                 - \int _{0}^x  \left( \dot{\phi} _1 (t,y) ^2 + \phi ^{\prime} (t,y) ^2 \right) \nu (y) \mu (y) dy, \quad \text{for} \; \delta (t,0) = 0 \\
                                  \int _{x}^{\pi /2}  \left( \dot{\phi} _1 (t,y) ^2 + \phi ^{\prime} (t,y) ^2 \right) \nu (y) \mu (y) dy, \quad \text{for} \; \delta (t,\pi / 2) = 0.\label{eq:delta2} \\
                                 \end{cases}
\end{eqnarray}
Here $\mu (x) = \text{tan}(x)^{d-1}$ and $\nu (x)= \frac{\text{sin}(x) \text{cos}(x)}{\text{tan}(x)^{d-1}}$. 

The first non--trivial dynamics appear at the $\mathcal{O}(\epsilon ^3)$ order, namely in the equation for $\phi _3$ in the {\it back--reacted} background. Here we will omit the details of the derivation of this equation and we will refer the reader to the numerous works where has already been presented\cite{BizRos11,CraEvn14,BalBuc14}. We will only mention, that to this order the field is expanded as:
\begin{eqnarray}
\phi _3 (t,x) = \sum _j c_{j}^{(3)}(t) e_j (x),
\end{eqnarray}
and the equation of motion for $\phi _3$ results to an infinite set of decoupled driven harmonic oscillators for the coefficients $c_j ^{(3)}(t)$. However, due to the highly commensurate spectrum of $AdS_{d+1}$, numerous resonances appear resulting in a secular growth of these coefficients at the time scale $t \sim \epsilon ^{-2}$ rendering this {\it naive} perturbation expansion invalid. A refined perturbation theory, known as the {\it Two Time Framework}\cite{BalBuc14,CraEvn14} consists of defining a {\it slow time}\footnote{The slow time variable $\tau = \epsilon ^2 t$ characterizes the time scale of the energy transfer between the normal modes while $t$ characterizes the oscillations of these normal modes.} $\tau = \epsilon ^2 t$ and allow the fields in eq.(\ref{eq:expansion}) depend on $\tau$ as well. The expansion would now be:
\begin{eqnarray}\label{eq:expansion2}
\phi = \sum_{k=0}^{\infty} \phi _{2k+1}(t,\tau,x) \epsilon ^{2k +1} , \quad A = 1 + \sum _{k=1}^{\infty} A_{2k}(t,\tau,x) \epsilon ^{2k }, \quad \delta = \sum _{k=1}^{\infty} \delta _{2k}(t,\tau,x) \epsilon ^{2k }. \nonumber \\
\end{eqnarray}
Now the resonances are entirely captured by the {\it slow time} evolution of the coefficients\footnote{For a detailed treatment of how these equations are obtained and for explicit expressions of the interaction coefficients, we refer the reader to the original papers\cite{BalBuc14,CraEvn14,CraEvn14a,GreMai15}.} in the expansion of $\phi _{1}$, eq.(\ref{eq:phi1expr}):
\begin{eqnarray}\label{eq:coeffevolution}
2i\omega _j \frac{d \alpha _j}{d \tau}  & = & T_j \abs{\alpha _j}^2 \alpha _j + \sum_{i \neq j} R_{ij} \abs{\alpha }_i ^2 \alpha _j+  \sum_{\substack{j+k = l+m \\ j \neq l, j\neq m}} S_{jklm} \alpha_k \alpha_l \alpha_m.
\end{eqnarray}
Using the {\it amplitude--phase} representation $\alpha _j (\tau) = A_j(\tau) e^{iB_j (\tau)}$ we can rewrite the above equation as:
\begin{eqnarray}
2 \omega _{j} \frac{d A_j}{d \tau} & = & \sum_{\substack{j+k = l+m \\ j \neq l, j\neq m}} S_{jklm} A_k A_l A_m \ \text{sin} \left( B_j + B_k - B_l - B_m \right)\label{eq:amplitudes} \\
2 \omega _j \frac{d B_j}{d \tau} & = & T_j A_j ^2 + \sum_{i \neq j} R_{ij}A_i ^2 + A_j ^{-1} \sum_{\substack{j+k = l+m \\ j \neq l, j\neq m}} S_{jklm} A_k A_l A_m \ \text{cos} \left( B_j + B_k - B_l - B_m \right). \label{eq:phases}  \nonumber \\
\end{eqnarray}


\section{Comparing the two gauges}
\label{sec-gauges}

In this section we compare the results in the two gauges. We will show that within the validity of TTF, they indeed describe the same physical evolution. The relation between the two gauges has also been studied in \cite{CraEvn14a} and some of the results can be found there as well. We will follow similar notations, but our attention lies on oscillating singularities that occur in one gauge and not the other. With some extra care we show what goes wrong as TTF breaks down when such a singularity develops in the central gauge.

The gauge choice should not affect any physical quantities. However, the two different gauges do lead to two different sets of differential equations, which were numerically evaluated to very different results. In\cite{BizMal15} the case of the two--mode equal energy data in $AdS_5$ was studied and an oscillating singularity was reported. Namely, the derivatives of the phases blow up. In\cite{Deppe16} it was shown that this singularity does not appear in the boundary gauge and therefore the singular behaviour of the system might be only an artefact of the gauge choice. 

On top of just numerical results, one can also see this difference from the asymptotic scaling of the $R_{ij}$ coefficients as was first suggested in\cite{CraEvn15a}. It was shown that for $AdS_5$ the $R_{ij}$ coefficients scale in the central gauge as $R_{ij}^{CG} \sim i^3 j^2$ and therefore, for a {\it power-law} spectrum $A_n \sim n^{-2}$ as observed in \cite{BizMal15}, the sum in the second term of eq.~(\ref{eq:phases}) diverges logarithmically. On the other hand, the asymptotic scaling of these coefficients in the boundary gauge was shown to be $R_{ij}^{BG} \sim i^2 j^2 $, thus although the evolution leads to the same power-law spectrum the same sum converges. One can check that the rest of the sums do not diverge. 

Despite this apparent difference, these results do not contradict each other. The oscillating singularity observed in\cite{BizMal15}, combined with the absence of that in \cite{Deppe16}, has an obvious physical meaning. It implies an infinite gravitational redshift between the boundary and the center of the spacetime.

From the metric (\ref{eq:aAdSmetric}), one can see that the two gauge choices are related as:
\begin{eqnarray}
dt_{BG} = e^{-\delta \left(t_{CG},\frac{\pi}{2} \right)} dt_{CG}.
\end{eqnarray}
Integrating and keeping terms only up to order  $\mathcal{O}(\epsilon^2)$ we get :
\begin{eqnarray}
t_{BG} & = & t_{CG} - \epsilon ^2  \int _0 ^{t_{CG}} dt \delta _2 (t,\tau,0) + \mathcal{O}(\epsilon ^4)
\end{eqnarray}
Neglecting terms that oscillate in the fast time scale $t$, we can approximate $\delta_2 (t,\tau,0)$ by the {\it time averaged} quantity $\delta _2 (\tau, 0)$. For completeness we will present the computation of this quantity in section (\ref{sec:redshift}). We then get:
\begin{eqnarray}\label{eq:timerelation}
 t_{BG} & \approx & t_{CG} - \epsilon ^2 \int _0 ^{t_{CG}} dt  \delta_2 (\tau,0) + \mathcal{O}(\epsilon ^4) \nonumber \\
           & = & t_{CG} + 2 \epsilon ^2 \int _0^{t_{CG}} dt \sum _{j} \left( A_{jj} + \omega ^2 _j V_{jj} \right) A_j ^2 + \mathcal{O}(\epsilon ^4),
\end{eqnarray}
Now, using the fact that the field $\phi (t,\tau,x)$ transforms as a scalar under such a {\it gauge transformation} one can derive the relation for the complex coefficients $\alpha _j (\tau)$ in the two gauges from eq.(\ref{eq:phi1expr}):
\begin{eqnarray}
\phi^{CG}(t_{CG}) & = & \phi ^{BG} (t_{BG}) \Rightarrow \nonumber \\
\alpha _j ^{CG} \left( \tau _{CG} \right)e^{i \omega _j t_{CG}} & = & \alpha _j ^{BG} \left( \tau _{BG} \right)e^{i \omega _j t_{BG}}
\end{eqnarray}
The relation of the slow time in the gauges is obtained simply by multiplying eq.(\ref{eq:timerelation}) by $\epsilon ^2$ to obtain
\begin{equation}
\tau_{BG} = \tau_{CG} + 2 \epsilon^2 \int_0^{\tau_{CG}} d\tau  \sum _{j} \left( A_{jj} + \omega ^2 _j V_{jj} \right) A_j ^2 + \mathcal{O}(\epsilon^4)
\end{equation}
 Substituting in the right hand side of the above equation, Taylor expanding and neglecting terms that are of order $\mathcal{O}(\epsilon ^2)$ we obtain:
\begin{eqnarray}
 \alpha _j ^{CG} \left( \tau _{CG} \right)e^{i \omega _j t_{CG}} & \approx & \left[ \alpha _j ^{BG} \left(\tau_{CG} \right) + \epsilon ^2 \dot{\alpha} ^{BG} \left( \tau_{CG} \right) \int _0 ^{\tau_{CG}} \delta _2 d\tau \right] \exp\left({i \omega _j t_{CG}} + {i \omega _j   \int_0^{\tau_{CG}} \delta _2 d\tau }\right) \nonumber \\
 \end{eqnarray}
Therefore, we find that the complex coefficients in the two gauges are related by:
\begin{eqnarray}\label{eq:coeffrelation}
\alpha _j ^{CG} \left( \tau  \right) & = &  \alpha _j ^{BG} \left( \tau  \right) \exp \left({i \omega _j  \int _0 ^{\tau} \delta _2 (\tau' ,0) d \tau'  } \right)+ \mathcal{O}(\epsilon ^2)
\end{eqnarray}
as is also explained in \cite{CraEvn14a}.
This result can also be expressed in the {\it amplitude--phase} representation, yielding:
\begin{eqnarray}
A_j ^{CG} (\tau) & = & A_j ^{BG} (\tau)\label{eq:amplituderelation} \\
B_{j} ^{CG} (\tau) & = &  B_{j} ^{BG} (\tau) - \omega _j \int ^{\tau}_{0} d \tau ' \sum _{i} \left( A_{ii} + \omega _i ^2 V_{ii} \right) A_{i} ^2(\tau ') \label{eq:phasediff}
\end{eqnarray}
That the amplitudes and the phases are related as above can be directly checked by applying eq.(\ref{eq:coeffrelation}) to the corresponding evolution equation in the two gauges, eq.(\ref{eq:coeffevolution}), and recalling that the difference is entirely contained in the coefficients\cite{CraEvn14a}:
\begin{eqnarray}
T_j^{BG} & = & T_j ^{CG} + \omega _j ^2 \left( A_{jj} + \omega _j ^2 V_{jj}  \right),\label{eq:gaugedifference1} \\
R_{ij}^{BG} & = & R_{ij} ^{CG} + \omega _j ^2 \left( A_{ii}  + \omega _i ^2 V_{ii}  \right)\label{eq:gaugedifference2}.
\end{eqnarray}

In\cite{FreYan15} it was shown that a large geometric back--reaction is related to the amplitude spectra and the coherence of the phases, where a phase--coherent cascade is defined by a spectrum of phases that (for large $j$) is linear in the mode number $j$:
\begin{eqnarray} \label{eq:phasecoherence}
B_j (\tau) = \gamma (\tau) j + \delta (\tau) + \dots,
\end{eqnarray}
This is an {\it asymptotic}\footnote{Recall also that asymptotically holds $\omega _j \approx j$, a fact that we use in eq.(\ref{eq:phaselinearity}).} statement and the ellipsis represent terms that are subleading in $j$. The reader should be aware here that the function $\delta (\tau)$ in the above equation is not the same function appearing in eq.~(\ref{eq:aAdSmetric}). From eq.(\ref{eq:amplituderelation}) we see that the evolution of the amplitudes is not affected by the choice of the gauge so what remains is to show that phase coherence is also unaffected and hence the physical conclusions will be independent of the choice of the gauge. Starting from eq.(\ref{eq:phasecoherence}) for the central gauge we have:
\begin{eqnarray}
B_j ^{CG} (\tau) & \approx & \gamma ^{CG} (\tau) j + \delta ^{CG} (\tau),
\end{eqnarray}
and applying eq.(\ref{eq:phasediff}) we can obtain the corresponding expression for the boundary gauge. This reads:
\begin{eqnarray}\label{eq:phaselinearity}
& B_j ^{BG} (\tau) & - \omega _j \int ^{\tau}_{0} d \tau ' \sum _{i} \left( A_{ii} + \omega _i ^2 V_{ii} \right) A_{i} ^2(\tau ') \approx  \gamma ^{CG} (\tau) j + \delta ^{CG} (\tau) \Rightarrow \nonumber \\ 
& B_j ^{BG}  (\tau) & \approx  \left(  \gamma ^{CG} (\tau) + \int ^{\tau}_{0} d \tau ' \sum _{i} \left( A_{ii} + \omega _i ^2 V_{ii} \right) A_{i} ^2(\tau ')  \right) j + \delta ^{CG} (\tau).\nonumber \\                    
\end{eqnarray}
We see that the phase spectrum in the boundary gauge takes the form of eq.(\ref{eq:phasecoherence}) :
\begin{eqnarray}
B_j ^{BG} (\tau) & \approx & \gamma ^{BG} (\tau) j + \delta ^{BG}(\tau), 
\end{eqnarray}
with the functions $\gamma (\tau)$ and $\delta (\tau)$ in the two gauges being related as:
\begin{eqnarray}
\gamma ^{BG} (\tau)  & = &  \gamma ^{CG} (\tau) + \int ^{\tau}_{0} d \tau ' \sum _{i} \left( A_{ii} + \omega _i ^2 V_{ii} \right) A_{i} ^2(\tau ')   \\
\delta ^{BG} (\tau) & = & \delta ^{CG} (\tau)
\end{eqnarray}


\subsection{The oscillating singularity as an infinite gravitational redshift} 
\label{sec:redshift}

Having clarified that physical conclusions can not be affected by the choice of the gauge, the next step is to reconcile the two different numerical results in the two gauges. In this section we will argue that the fact that $\dot{B}_j$ diverges in the one gauge and not in the other can be interpreted as an {\bf infinite gravitational redshift} between the boundary and the center of the spacetime. Recall that the gravitational redshift between a source and an observer is given by the formula:
\begin{eqnarray}
1 + z = \sqrt{\frac{g_{tt}(obs)}{g_{tt}(source)}}.
\end{eqnarray}
We can compute this quantity in one of the two gauges. Let us choose the nonsingular {\it boundary gauge} and compute the redshift between the boundary ($x=\pi / 2$) and the center ($x=0$) of the spacetime. Using the metric (\ref{eq:aAdSmetric}), the normalization $\delta (t,\pi /2 ) = 0$ and keeping terms only up to the order of $\mathcal{O}(\epsilon ^2)$, the quantity under the square root reads:
\begin{eqnarray}\label{eq:redshift}
\frac{g_{tt}(t,0)}{g_{tt}(t,\pi / 2)} & \sim & 1 -\epsilon ^2 \delta _2(t,\tau,0) + \mathcal{O}(\epsilon ^4).
\end{eqnarray}
The expression for $\delta _2 (t,\tau,0)$, eq.(\ref{eq:delta2}), yields\footnote{For ease of notation we have omitted to write explicitly the slow time dependence in some cases, but it is implicitly assumed.}:
\begin{eqnarray}
\delta _2 (t,\tau,0) & = & \int_0 ^{\pi / 2 }  \left( \dot{\phi} _1 (t,x) ^2 + \phi _1 ^{\prime} (t,x) ^2 \right) \mu(x) \nu (x) dx\nonumber \\
                      & = & \int_0 ^{\pi / 2 }  \sum_{ij} \left( \dot{c}^{(1)}_i (t) \dot{c}^{(1)}_j (t) e_i(x) e_j(x) + c ^{(1)}_i (t) c ^{(1)}_j (t) e_i (x) e_j (x) \right) \mu(x) \nu (x) dx  \nonumber \\
                      & = & \sum_{ij} \left( \dot{c}^{(1)}_i  \dot{c}^{(1)}_j  V_{ij} + c^{(1)}_i c^{(1)}_j A_{ij} \right).
\end{eqnarray}
To go to the second line, we simply used the expansion in eigenmodes $\phi _1 (t,x) = \sum _j c_j ^{(1)}(t)e_j (x)$ and in the third line we defined the interaction coefficients:
\begin{eqnarray}
A_{ij} & \equiv & \int _{0}^{\pi /2} e_i ^{\prime}(x) e_j ^{\prime}(x) \mu (x) \nu (x) dx \\
V_{ij} & \equiv & \int _{0}^{\pi /2} e_i (x) e_j (x) \mu (x) \nu (x) dx.
\end{eqnarray}
The expansion coefficients $c_j ^{(1)}$ are related to the complex coefficients $ \alpha _j $ as\footnote{As we have stated below eq.(\ref{eq:coeffevolution}) these are also related to $A_j$ and $B_j$ coefficients as: $A_j = \abs{\alpha _j}$ and $B_j = Arg(\alpha _j)$.}:
\begin{eqnarray}
c^{(1)}_j & = & \alpha _j e^{i \omega _j t} + \bar{\alpha} _j e^{-i \omega _j t}\label{eq:complexcoeff} \\
\frac{d c^{(1)}_j}{dt} & = & i \omega _j \left( \alpha _j e^{i \omega _j t} - \bar{\alpha} _j e^{-i \omega _j t} \right).
\end{eqnarray}
Substituting eq.(\ref{eq:complexcoeff}) in the above expression for $\delta _2 (t,\tau,0)$ we will get several terms of the form $e^{i \Omega t}$, where $\Omega = \omega _i \pm \omega _j$. Keeping only terms with $\Omega = 0$, the so called {\it resonant} terms\footnote{These are are the terms that are proportional to $e^{\pm i (\omega _i - \omega_j)}\delta_{ij}$. This procedure is equivalent to time--averaging over the {\it fast time} $t$.}, we finally obtain the following expression:
\begin{eqnarray}
\delta _2 (t,\tau,0) \approx  2 \sum_{i} \left( A_{ii} + \omega _i ^2 V_{ii} \right) A^{2}_i (\tau) \equiv \delta _2 (\tau,0) .
\end{eqnarray}
By differentiating eq.(\ref{eq:phasediff}), we can see that this quantity, the {\it time--averaged} $\delta_2$, which was first mentioned in eq.(\ref{eq:timerelation}), is precisely the difference of the slow time derivatives of the phases in the two gauges. Therefore, by comparing the results in the boundary and the central time gauge we can draw conclusions about geometric quantities, and in particular the gravitational redshift. In the case of interest, where the derivatives of the phases diverge in one gauge but not in the other, one concludes that $\delta _2 (\tau,0)$ diverges, and so does the redshift, eq.(\ref{eq:redshift}). This large back-reaction in turn implies the breakdown of linearized gravity. 
On the other hand if the derivatives are finite in both gauges there is no divergence, while the case is not clear if an oscillating singularity appears both in the boundary as well as in the central gauge. In that case $\delta _2 (\tau,0)$ could be either finite or infinite.

\section{Conclusions}

In this manuscript we presented an explicit derivation on the anticipated fact that physical results can not be affected by the different gauge choices. We demonstrated that gauge-invariant quantities are related to the amplitude spectrum and the coherence of the phases in the TTF solution, and both properties are unaffected by the gauge choice. This result holds even when the difference between the two gauges diverges. Furthermore we established that the oscillating singularity observed in\cite{BizMal15} is indeed a physical singularity, by showing that is related to an infinite redshift between the boundary and the center of the spacetime.

This means that the breakdown of the TTF observed in \cite{BizMal15} is due to large gravitational effects which lead to the breakdown of the weak gravity approximation. Such a conclusion cannot be deduced by the observed singularity in the central time gauge alone. In that case is not clear whether the breakdown of the perturbation theory is caused by strong gravity or by the breakdown of other approximations. Therefore, with our analysis we establish that the singular solution is a genuine singular solution of the gravitational problem. Due to the scaling symmetry of the TTF system the solution will survive in the $\epsilon \rightarrow 0$ limit, and thus provide a way to address the phase space of initial conditions in this limit.

An interesting thing to point out here is that for this conclusion we need to compare the derivatives of the phases in the two gauges. Therefore, the fact that in higher dimensions  a discrepancy between the two gauges has not been observed\cite{Deppe16} is rather intriguing. However since in both gauges an oscillating singularity was observed, and actually in the central time gauge this divergence was more prominent than in the boundary time gauge, it might still signal a diverging redshift, since these results are compatible with a diverging $\delta _2 (\tau,0)$, as we explained in Section \ref{sec:redshift}.

\acknowledgments

We thank  Andrzej Rostworowski and Matthew Lippert for useful discussions. This work is part of the $\Delta$-ITP consortium and also supported in part by the Foundation for Fundamental Research on Matter (FOM), both are parts of the Netherlands Organisation for Scientific Research (NWO) that is funded by the Dutch Ministry of Education, Culture and Science (OCW). F.D. is supported by GRAPPA PhD Fellowship.

\appendix




\bibliographystyle{utcaps}
\bibliography{all_active}

\providecommand{\href}[2]{#2}\begingroup\raggedright\begin{thebibliography}{10}

\bibitem{BizMal15}
P.~Bizon, M.~Maliborski, and A.~Rostworowski, ``{Resonant dynamics and the
  instability of anti-de Sitter spacetime},''
\href{http://arxiv.org/abs/1506.03519}{{\tt arXiv:1506.03519 [gr-qc]}}.

\bibitem{Deppe16}
N.~Deppe, ``{On the stability of anti-e Sitter spacetime},''
\href{http://arxiv.org/abs/1606.02712}{{\tt arXiv:1606.02712 [gr-qc]}}.

\bibitem{CraEvn15a}
B.~Craps, O.~Evnin, and J.~Vanhoof, ``{Ultraviolet asymptotics and singular
  dynamics of AdS perturbations},''
  \href{http://dx.doi.org/10.1007/JHEP10(2015)079}{{\em JHEP} {\bf 10} (2015)
  079},
\href{http://arxiv.org/abs/1508.04943}{{\tt arXiv:1508.04943 [gr-qc]}}.

\bibitem{BizRos11}
P.~Bizon and A.~Rostworowski, ``{On weakly turbulent instability of anti-de
  Sitter space},'' \href{http://dx.doi.org/10.1103/PhysRevLett.107.031102}{{\em
  Phys.Rev.Lett.} {\bf 107} (2011)  031102},
\href{http://arxiv.org/abs/1104.3702}{{\tt arXiv:1104.3702 [gr-qc]}}.

\bibitem{MalRos13a}
M.~Maliborski and A.~Rostworowski, ``{Time-Periodic Solutions in an Einstein
  AdS--Massless-Scalar-Field System},''
  \href{http://dx.doi.org/10.1103/PhysRevLett.111.051102}{{\em Phys.Rev.Lett.}
  {\bf 111} (2013) no.~5, 051102},
\href{http://arxiv.org/abs/1303.3186}{{\tt arXiv:1303.3186 [gr-qc]}}.

\bibitem{GreMai15}
S.~R. Green, A.~Maillard, L.~Lehner, and S.~L. Liebling, ``{Islands of
  stability and recurrence times in AdS},''
  \href{http://dx.doi.org/10.1103/PhysRevD.92.084001}{{\em Phys. Rev.} {\bf
  D92} (2015) no.~8, 084001},
\href{http://arxiv.org/abs/1507.08261}{{\tt arXiv:1507.08261 [gr-qc]}}.

\bibitem{CraEvn14}
B.~Craps, O.~Evnin, and J.~Vanhoof, ``{Renormalization group, secular term
  resummation and AdS (in)stability},''
\href{http://arxiv.org/abs/1407.6273}{{\tt arXiv:1407.6273 [gr-qc]}}.

\bibitem{BucGre14}
A.~Buchel, S.~R. Green, L.~Lehner, and S.~L. Liebling, ``{Conserved quantities
  and dual turbulent cascades in Anti-de Sitter spacetime},''
\href{http://arxiv.org/abs/1412.4761}{{\tt arXiv:1412.4761 [gr-qc]}}.

\bibitem{DiaHor11}
O.~J. Dias, G.~T. Horowitz, and J.~E. Santos, ``{Gravitational Turbulent
  Instability of Anti-de Sitter Space},''
  \href{http://dx.doi.org/10.1088/0264-9381/29/19/194002}{{\em
  Class.Quant.Grav.} {\bf 29} (2012)  194002},
\href{http://arxiv.org/abs/1109.1825}{{\tt arXiv:1109.1825 [hep-th]}}.

\bibitem{DiaHor12}
O.~J. Dias, G.~T. Horowitz, D.~Marolf, and J.~E. Santos, ``{On the Nonlinear
  Stability of Asymptotically Anti-de Sitter Solutions},''
  \href{http://dx.doi.org/10.1088/0264-9381/29/23/235019}{{\em
  Class.Quant.Grav.} {\bf 29} (2012)  235019},
\href{http://arxiv.org/abs/1208.5772}{{\tt arXiv:1208.5772 [gr-qc]}}.

\bibitem{DiaSan16}
O.~J. Dias and J.~E. Santos, ``{AdS nonlinear instability: moving beyond
  spherical symmetry},''
\href{http://arxiv.org/abs/1602.03890}{{\tt arXiv:1602.03890 [hep-th]}}.

\bibitem{DFLY14}
F.~V. Dimitrakopoulos, B.~Freivogel, M.~Lippert, and I.-S. Yang, ``{Instability
  corners in AdS space},''
\href{http://arxiv.org/abs/1410.1880}{{\tt arXiv:1410.1880 [hep-th]}}.

\bibitem{DimYan15}
F.~Dimitrakopoulos and I.-S. Yang, ``{Occasionally Extended Validity of
  Perturbation Theory: Persistence of AdS Stability Islands},''
\href{http://arxiv.org/abs/1507.02684}{{\tt arXiv:1507.02684 [hep-th]}}.

\bibitem{FreYan15}
B.~Freivogel and I.-S. Yang, ``Coherent Cascade Conjecture for Collapsing
  Solutions in Global AdS,'' {\em Phys. Rev. D} {\bf 93} (2016)  103007,
\href{http://arxiv.org/abs/hep-th/1512.04383}{{\tt hep-th/1512.04383}}.

\bibitem{OliSop16}
D.~S. Olivan and C.~F. Sopuerta, ``{Moving Closer to the Collapse of a Massless
  Scalar Field in Spherically Symmetric Anti-de Sitter Spacetimes},''
\href{http://arxiv.org/abs/1603.03613}{{\tt 1603.03613}}.

\bibitem{BalBuc14}
V.~Balasubramanian, A.~Buchel, S.~R. Green, L.~Lehner, and S.~L. Liebling,
  ``{Holographic Thermalization, Stability of AdS, and the FPU Paradox},''
\href{http://arxiv.org/abs/1403.6471}{{\tt arXiv:1403.6471 [hep-th]}}.

\bibitem{CraEvn14a}
B.~Craps, O.~Evnin, and J.~Vanhoof, ``{Renormalization, averaging, conservation
  laws and AdS (in)stability},''
\href{http://arxiv.org/abs/1412.3249}{{\tt arXiv:1412.3249 [gr-qc]}}.

\end{thebibliography}\endgroup

\end{document}